# Giant Cotton–Mouton effect of suspensions of iron nanorods

Shu-guo Lei and Cheng-ping Huang*

*Department of physics, Nanjing Tech University, Nanjing 211816, China*

**Abstract**

The Cotton-Mouton (CM) effect, referring to linear birefringence induced by a magnetic field, is usually very weak in natural materials. We propose theoretically that a giant CM effect may be achieved in the THz region with the suspension of iron nanorods. The unusual effect stems from the dual nature of the iron nanorods, which exhibit both ferromagnetic and plasmonic characteristics. Our results suggest that, with an ultralow magnetic field ~20 mT, a large linear birefringence up to ~0.10 can be realized. The CM coefficient attains $2.2\times10^7$ $T^{-2}m^{-1}$, four orders of magnitude larger than that obtained currently in optical range. The results may advance the study and applications of the THz CM effect in artificial magneto-optic materials.

*cphuang@njtech.edu.cn

The interaction of light with matter in the presence of a magnetic field gives rise to a fascinating class of phenomena known as magneto-optic effects [1, 2], which provide powerful probes of material properties and enable important photonic applications. A well-known example is the Faraday effect, where the polarization plane of linearly polarized light can rotate as it propagates through a material parallel to an applied magnetic field [3]. This non-reciprocal effect, arising from magnetic circular birefringence, is crucial for constructing optical isolators in laser systems and fiber-optic communications. Another example is the Cotton-Mouton (CM) effect, initially proposed by A. Cotton and H. Mouton in 1905, which appears when light propagates perpendicular to the magnetic field [4]. Because of the field-induced anisotropy, a linear birefringence effect proportional to the square of the magnetic field will be resulted. However, compared with the electro-optic effect of crystals, the linear birefringence of the CM effect is usually very weak for liquid or gas ($\Delta n=10^{-16} \sim 10^{-11}$) [5, 6], prohibiting its wide theoretical interests and practical applications. Thus, realizing the CM effect with low operating field and large linear birefringence becomes a challenge.

To address the issue, the magnetic fluids consisting of the spherical or ellipsoidal magnetic nanoparticles (~10 nm) in a liquid host medium have been studied [7-9]. But due to the small sizes and negligible shape anisotropy, the magnetic and anisotropic optical responses of these nanoparticles are relatively weak, corresponding to a minor magnetic linear birefringence of $\Delta n=10^{-6} \sim 10^{-5}$ [7-9]. Besides, the magnetic fluids exhibit a strong absorption in the visible region, which is harmful for the optical devices [8]. Recently, the suspensions of 2D crystals, e.g., the cobalt-doped titanium oxide, have been shown to own a large CM effect, where a higher induced birefringence of $\Delta n=2\times 10^{-4}$ (at the magnetic field 0.8 T) and a weak absorption can be achieved simultaneously [10]. Moreover, similar optical effects have been extended successfully into the deep ultraviolet range, thus demonstrating great potential of the 2D material system for the future applications [11, 12].

In this paper, a giant CM effect has been suggested theoretically in the opposite low frequency (THz) range, employing the suspensions of iron nanorods. The unusual effect is due to the ferromagnetic and plasmonic properties of the iron nanorods. On one hand, the ferromagnetic iron nanorod can exhibit a large magnetic dipole moment, which enables the orientation of nanorods with a low magnetic field. On the other hand, the metallic iron nanorods can support the plasmonic antenna resonance, leading to a strong optical response in the THz band. These characteristics make a significant difference from the magnetic fluids or 2D material systems [7-12]. Our results show that, with an ultralow magnetic field of ~20 mT, a large linear birefringence of ~0.10 can be achieved in the THz region. The results are valuable for designing novel THz magneto-optic materials and devices.

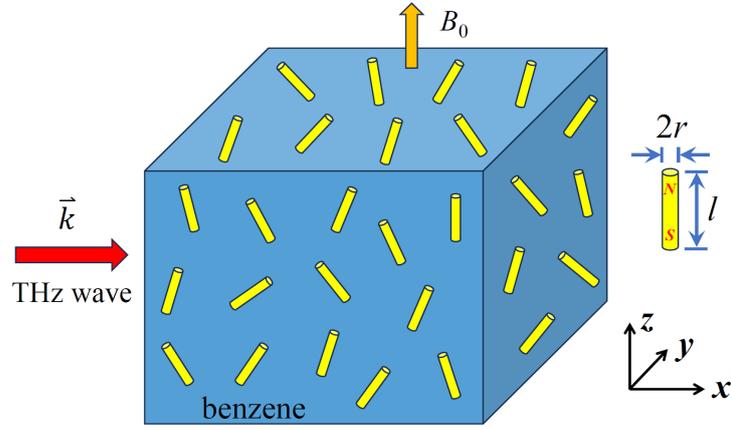

*Fig. 1 Schematic view of the system: the ferromagnetic and plasmonic iron nanorods are randomly dispersed in the non-polar liquid medium (benzene). A THz wave propagates along the x axis and the static magnetic field $B_0$ is exerted along the z axis.*

To study the CM effect, Fig. 1 presents the schematic view of the system. In the suspension, the ferromagnetic and plasmonic iron nanorods with the length $l$ and radius $r$ ($l \gg r$) are well separated and randomly dispersed in the liquid host medium. The temperature of the suspension is $T$ and the number density of the nanorods is $N$. To reduce the dissipation, the non-polar liquid, benzene, is utilized as the host medium, which exhibits negligible dispersion and absorption in the THz band ($\varepsilon_d = 2.25$) [13]. The THz wave is propagating in the suspension along the $x$ axis and a static magnetic field $B_0$ is exerted on the system along the $z$ axis. Assuming that the radius $r$ is smaller than the characteristic domain size (~50 nm) [14, 15], the iron nanorod thus becomes a ferromagnetic single-domain structure (in practice, the iron nanorods can be encapsulated within a silica layer and subsequently functionalized with silane coupling agents to prevent both corrosion and aggregation [16, 17]). The plasmonic nature of the iron can be described with the Drude model, where the plasma frequency is $\omega_p$=6.22×10$^{15}$ rad/s and the electron collision frequency is $\gamma$=2.77×10$^{13}$ Hz [18].

The ferromagnetic single-domain iron nanorod owns an inherent magnetic moment (a macroscopic spin vector) $m_0 = M_r \Delta v$, where $M_r$ is the remanent magnetization and $\Delta v = \pi r^2 l$ is the volume of the nanorod. Due to the high shape anisotropy, this magnetic moment is very robust and orients along the long nanorod axis, i.e., the easy magnetization axis [15]. Under the applied magnetic field, the iron nanorods will tend to be aligned, resulting in a net magnetization in $z$ direction. The orientation of the nanorods is described by the Boltzmann distribution $f(\theta) = e^{-U/k_B T} = e^{\beta \cos \theta}$, where $k_B$ is the Boltzmann constant, $\theta$ is the angle between the nanorod and the $z$ axis, and $U = -m_0 B_{0,eff} \cos \theta$ is the orientation energy of the nanorod. Thus, the orientation distribution of the nanorods is governed by

$$\beta = \frac{m_0}{k_B T} B_{0,eff}. \tag{1}$$

Here, $B_{0,eff}$ is the effective magnetic field acting on the nanorod, which includes both the external field $B_0$ and the interactions of the iron nanorods.

For the low nanorod concentration and random spatial distribution, the effective magnetic field acting on the nanorods may be determined approximately in the framework of the Lorentz sphere model [19]. Following the method, the effective field can be expressed as $B_{0,eff} = B_0 + \mu_0 M_{0,z}/3$, where $M_{0,z} = N \langle m_z \rangle$ represents the macroscopic magnetization intensity, $\langle m_z \rangle = m_0 X_1$ is the average value (z-component) of the magnetic moments, and $X_1 = \langle \cos \theta \rangle$ is the Langevin function:

$$X_1 = \frac{\int_{-1}^{1} u e^{\beta u} du}{\int_{-1}^{1} e^{\beta u} du} = \coth\beta - \frac{1}{\beta}. \tag{2}$$

Thus, the effective magnetic field can be deduced as $B_{0,eff} = B_0 + \mu_0 m_0 N X_1 / 3$. With the above results, the quantitative relationship between the parameter $\beta$ and the external magnetic field $B_0$ can be determined.

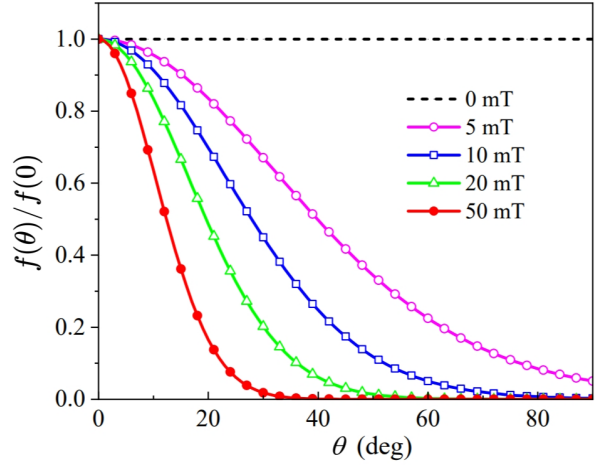

*Fig. 2 Normalized probability distribution function versus the orientation angle of the iron nanorod. Here, l=1 μm, r=10 nm, T=300 K, ρ=2 mg/ml, and the magnetic field is chosen as $B_0$=0, 5, 10, 20, and 50 mT, respectively.*

To evaluate the alignment ability of the nanorods by the magnetic field, Fig. 2 plots the normalized distribution function $f(\theta)/f(0)$ as a function of $\theta$ for various magnetic field. Here, the sizes of the nanorods are set as $l$=1 μm and $r$=10 nm, the temperature of suspension is $T$=300 K, the concentration of the nanorods is $\rho$=2 mg/ml ($N$=8.1×10$^{11}$/ml with a volume fraction of 0.025%), and the remanent magnetization of the iron nanorods is $M_r$=7.87×10$^3$ A/m (or 1.0 emu/g) [20, 21] (the parameters will be used throughout this work). One can see that, without the magnetic field, the orientation distribution is uniform in the space. Once a magnetic field is present, the distribution function decays monotonically with $\theta$. The larger the magnetic field, the rapid the decaying rate. Strikingly, an ultralow magnetic field of tens of mT can induce a pronounced anisotropy in the orientation distribution, progressively aligning the nanorods towards the field direction. The effect is correlated with the remanent magnetization and larger volume of the nanorods, yielding a considerable magnetic moment and a weak orientation field. In contrast, the 2D nanosheet with paramagnetism and nanoscale thickness owns small induced magnetic moment, thus requiring a high magnetic field (~0.8 T) [10-12]. The result provides the possibility for controlling the iron nanorods and generating the THz CM effect with an ultralow magnetic field.

The alignment of the iron nanorods by the magnetic field may induce a significant birefringence effect. Due to the excitation of the THz wave, an oscillating electric dipole moment can be induced in the nanorod, thus giving rise to the THz optical response of the suspension. In the long wavelength limit (the wavelength of the THz wave is much larger than the length of the nanorods), the suspension can be treated as a homogeneous medium with the effective permittivity [22]. The permittivity of the two orthogonal polarization states, corresponding to the wave field parallel ($E//z$) or perpendicular ($E \perp z$) to the static magnetic field, can be derived as

$$\varepsilon_{//} = \varepsilon_d + \frac{N\alpha X_2}{1 - N\alpha X_2/3\varepsilon_d},$$
$$\varepsilon_{\perp} = \varepsilon_d + \frac{N\alpha(1-X_2)}{2 - N\alpha(1-X_2)/3\varepsilon_d}. \tag{3}$$

Here, $\alpha$ is the THz optical polarizability of the iron nanorod and $X_2 = \langle \cos^2\theta \rangle$ is the generalized Langevin function, which can be expressed as

$$\alpha = \frac{Cl_{eff}^2/\varepsilon_0}{1-(\omega/\omega_0)^2 - i\eta(\omega/\omega_0)},$$
$$X_2 = \frac{\int_{-1}^{1} u^2 e^{\beta u} du}{\int_{-1}^{1} e^{\beta u} du} = 1 - \frac{2}{\beta \tanh\beta} + \frac{2}{\beta^2}, \tag{4}$$

where $C$ is the capacitance of the nanorod, $l_{eff} = 2l/\pi$ is the effective rod length, $\varepsilon_0$ is the vacuum permittivity, and $\omega_0$ is the resonance frequency of the nanorod [23].

One can see that the effective permittivity is strongly dependent on the THz optical polarizability $\alpha$. Due to the plasmonic nature, the iron nanorod will experience an antenna resonance at the frequency $\omega_0$. Near the resonance, numerous free electrons in the iron nanorod can displace oscillatorily at a long distance (~ the rod length $l$, which is of micrometer scale), thereby resulting in a large optical polarizability. By contrast, the 2D nanosheets employed in previous works are electric insulators [10-12], where the bounded electrons can only displace at the nanoscale, thus limiting the magnitude of the polarizability. In addition, the magnitude of $\alpha$ can be further boosted by increasing the rod length and accordingly lowering the resonance/working frequency. For the sizes used in this work, the iron nanorod has a resonance frequency of 26 THz. At the working frequency $f = 3$ THz (which is off resonance to suppress the absorption), for instance, the real and imaginary parts of $\alpha$ is $3.4 \times 10^{-19}$ and $6.1 \times 10^{-21}$ m$^3$, respectively. Compared with the molecules (~$10^{-29}$ m$^3$), the THz optical polarizability of the iron nanorods is enhanced by 10 orders of magnitude. This is responsible for the large modulation range of the refractive index.

The above results also suggest that, with the increase of the magnetic field $B_0$, the generalized Langevin function $X_2$, the effective permittivity, and the refractive index ($n_{//} = \sqrt{\varepsilon_{//}}$ and $n_\perp = \sqrt{\varepsilon_\perp}$) can be modified. To clearly see the effect, we consider the case of smaller value of $\beta$ (where $X_2 \approx 1/3 + 2\beta^2/45$ and $B_{0,eff} \approx B_0$). In this case, the effective permittivity can be rewritten as

$$\begin{aligned}\varepsilon_{//} &= a + 2bB_0^2, \\ \varepsilon_\perp &= a - bB_0^2,\end{aligned} \quad (5)$$

where

$$\begin{aligned}a &= \varepsilon_d + \frac{N\alpha/3}{1 - N\alpha/9\varepsilon_d}, \\ b &= \frac{N\alpha}{45}(\frac{m_0/k_BT}{1 - N\alpha/9\varepsilon_d})^2.\end{aligned} \quad (6)$$

Equation (5) shows that, for both polarization states, the effective permittivity exhibits a quadratic relationship with the magnetic field. However, the variation tendency is just opposite: one increases and the other decreases with the field. Thus, the suspension exhibits a birefringence effect, which amplifies with the magnetic field. With Eqs. (5) and $\Delta n = C\lambda B_0^2$, the THz CM coefficient can be derived approximately as

$$C \approx \frac{N\alpha}{30\sqrt{\varepsilon_d}\lambda}(\frac{m_0}{k_B T})^2, \qquad (7)$$

which is proportional to the concentration, the THz optical polarizability, and the square of the magnetic moment of the iron nanorods.

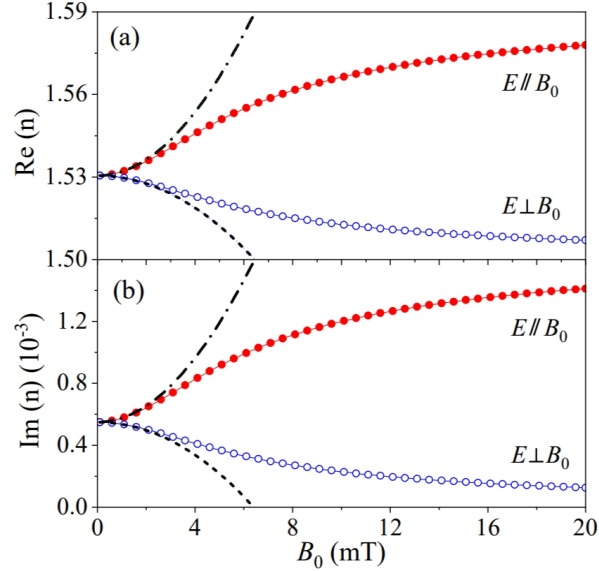

Fig. 3 Dependence of refractive index on the magnetic field: (a) the real part and (b) the imaginary part. The solid and open circles correspond to the exact results for $E//B_0$ and $E \perp B_0$, respectively (the dash lines represent the approximated results). Here, the frequency of the THz wave is $f=3$ THz.

To quantitatively study the CM effect, the dependence of refractive index on the applied magnetic field $B_0$ has been calculated using Eqs. (3) and (5) and the results are shown in Fig. 3 ($f=3$ THz). Here, the solid and open circles correspond to the exact solutions for the parallel ($E//B_0$) or perpendicular ($E \perp B_0$) polarization, respectively; the dash lines represent the approximated results, which are close to the exact one only at the weak magnetic field ($B_0<4$ mT). With the increase of $B_0$, a clear increase of real and imaginary parts of refractive index, Re($n_{//}$) and Im($n_{//}$), for the parallel polarization can be observed (the situation is just opposite for the perpendicular polarization). To be specific, when $B_0$ increases from 0 to 20 mT, Re($n_{//}$) grows gradually from 1.530 to 1.578, showing an increment of $\Delta n_{//}=0.048$. Correspondingly, Re($n_\perp$) decreases from 1.530 to 1.506, with a reduction of $\Delta n_\perp =-0.024$. When $B_0$=20 mT, a significant birefringence effect with $\Delta n=\text{Re}(n_{//}-n_\perp)=0.072$ can be achieved. With the further increase of $B_0$, the modulation will become saturated, as the nanorods have been sufficiently aligned by the field. On the other hand, when $B_0$ varies from 0 to 20 mT, Im($n_{//}$) increases from $5.50\times10^{-4}$ to $1.41\times10^{-3}$ and Im($n_\perp$) reduces from $5.50\times10^{-4}$ to $1.26\times10^{-4}$, respectively. The imaginary part of the refractive index is more than 3 orders of magnitude smaller than the real part, which is desirable for the THz wave propagation. The low operating field and high refractive-index modulation represent a giant CM effect achieved in the THz band. With Eq. (7), the THz CM coefficient has been determined as $C=2.2\times10^7$ T$^{-2}$m$^{-1}$, which is about 4 orders of magnitude larger than that achieved with the magnetic fluids or 2D crystals (~$10^3$ T$^{-2}$m$^{-1}$) [7-12].

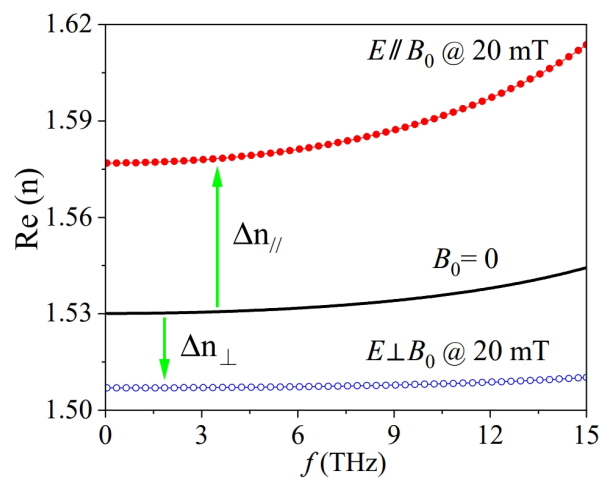

*Fig. 4 Dispersion effect (the real part of refractive index) of the iron nanorod suspension. The solid lines show the dispersion without magnetic field; the solid and open circles show the dispersion with E//B₀ and E ⊥B₀ (B₀=20 mT), respectively.*

As the THz optical polarizability $\alpha$ is a function of frequency, the refractive index of the suspension will exhibit an obvious dispersion effect. Figure 4 shows the real part of refractive index as a function of the THz wave frequency (the imaginary part is less than 0.015 and not shown here). Without the external magnetic field (the solid lines), Re(n) varies from 1.530 to 1.544 when the frequency $f$ increases from 0.1 to 15 THz. When a magnetic field $B_0$=20 mT is applied, the refractive index for the parallel polarization shows a larger dispersion effect (the solid circles), where Re($n_{//}$) increases from 1.577 to 1.614. On the contrary, the refractive index for the perpendicular polarization becomes almost dispersionless [Re($n_\perp$) is close to 1.510; the open circles]. Note that, at the frequency of 15 THz, a magnetic linear birefringence up to ~0.10 can be achieved, which is comparable with the birefringence effect of liquid crystals. The results show that the iron nanorod suspension can exhibit a significant CM effect and weak absorption over a wide THz frequency range.

In conclusion, a giant THz CM effect of the suspension of iron nanorods have been suggested. The dual nature, i.e., the ferromagnetic and plasmonic characteristics, of the iron nanorods provides a unique and highly efficient platform for manipulating the magneto-optic properties. We show that a large refractive-index modulation (up to 0.10) can be achieved with an ultralow working field (~20 mT), thus corresponding to a significant enhancement of the CM coefficient. The results are valuable for studying the THz CM effect in artificial magneto-optic materials and for designing high-performance THz devices [24-26].

This work was supported by the National Natural Science Foundation of China (Grant No. 12174193).